\documentclass[
reprint,superscriptaddress,amsmath,amssymb,aps,prx,twocolumn,nofootinbib]{revtex4-2}

\usepackage{graphicx}
\usepackage{dcolumn}
\usepackage{bm}
\usepackage{hyperref}
\hypersetup{colorlinks = true, linkcolor=blue, citecolor={blue}, urlcolor={blue}}
\usepackage{academicons}

\newcommand{\orcid}[1]{\href{https://orcid.org/#1}{\textcolor[HTML]{A6CE39}{\aiOrcid}}}
\usepackage{scalerel}
\usepackage{tikz}
\usetikzlibrary{svg.path}

\newcommand{\PRLsection}[1]{\textit{#1} ---}

\definecolor{orcidlogocol}{HTML}{A6CE39}
\tikzset{
    orcidlogo/.pic={
        \fill[orcidlogocol] svg{M256,128c0,70.7-57.3,128-128,128C57.3,256,0,198.7,0,128C0,57.3,57.3,0,128,0C198.7,0,256,57.3,256,128z};
        \fill[white] svg{M86.3,186.2H70.9V79.1h15.4v48.4V186.2z}
        svg{M108.9,79.1h41.6c39.6,0,57,28.3,57,53.6c0,27.5-21.5,53.6-56.8,53.6h-41.8V79.1z M124.3,172.4h24.5c34.9,0,42.9-26.5,42.9-39.7c0-21.5-13.7-39.7-43.7-39.7h-23.7V172.4z}
        svg{M88.7,56.8c0,5.5-4.5,10.1-10.1,10.1c-5.6,0-10.1-4.6-10.1-10.1c0-5.6,4.5-10.1,10.1-10.1C84.2,46.7,88.7,51.3,88.7,56.8z};
    }
}

\newcommand\orcidicon[1]{\href{https://orcid.org/#1}{\mbox{\scalerel*{
                \begin{tikzpicture}[yscale=-1,transform shape]
                \pic{orcidlogo};
                \end{tikzpicture}
            }{|}}}}

\usepackage[nameinlink]{cleveref}
\usepackage{siunitx}

\begin{document}

\title{Interplay between Nuclear Shell Structure and Pairing around Doubly Magic $^{132}$Sn}

\author{R.~Simpson\text{\orcidicon{0009-0002-7714-5649}}}
\email{ranes@triumf.ca}
\affiliation{TRIUMF, Vancouver, British Columbia, V6T 2A3, Canada}
\affiliation{Department of Physics \& Astronomy, University of British Columbia, Vancouver, British Columbia V6T 1Z1, Canada}

\author{G.~Palkanoglou\text{\orcidicon{0000-0001-8543-6097}}}
\email{gpalkanoglou@triumf.ca}
\affiliation{TRIUMF, Vancouver, British Columbia, V6T 2A3, Canada}

\author{E.C.~Brisley}
\affiliation{TRIUMF, Vancouver, British Columbia, V6T 2A3, Canada}
\affiliation{Department of Physics \& Astronomy, University of British Columbia, Vancouver, British Columbia V6T 1Z1, Canada}

\author{J.D.~Cardona}
\affiliation{TRIUMF, Vancouver, British Columbia, V6T 2A3, Canada}
\affiliation{Department of Physics and Astronomy, University of Manitoba, Winnipeg, Manitoba, R3T 2N2 Canada}

\author{A.~Czihaly}
\affiliation{TRIUMF, Vancouver, British Columbia, V6T 2A3, Canada}
\affiliation{Department of Physics and Astronomy, University of Victoria, Victoria, British Columbia, V8O 5C2, Canada}

\author{S.~Kakkar}
\affiliation{TRIUMF, Vancouver, British Columbia, V6T 2A3, Canada}
\affiliation{Department of Physics and Astronomy, University of Manitoba, Winnipeg, Manitoba, R3T 2N2 Canada}

\author{M.~Simonov}
\affiliation{II. Physikalisches Institut, Justus-Liebig-Universität, Gießen, 35392, Germany.}
\affiliation{Helmholtz Forschungsakademie Hessen für FAIR (HFHF), Campus Gießen, Gießen, 35392, Germany.}

\author{E.~Taylor}
\affiliation{TRIUMF, Vancouver, British Columbia, V6T 2A3, Canada}
\affiliation{Department of Physics and Astronomy, University of Western Ontario, London, Ontario, N6A 3K7 Canada}

\author{C.~Walls}
\affiliation{TRIUMF, Vancouver, British Columbia, V6T 2A3, Canada}
\affiliation{Department of Physics and Astronomy, University of Manitoba, Winnipeg, Manitoba, R3T 2N2 Canada}

\author{P.~Weligampola}
\affiliation{TRIUMF, Vancouver, British Columbia, V6T 2A3, Canada}
\affiliation{Department of Physics and Astronomy, University of Manitoba, Winnipeg, Manitoba, R3T 2N2 Canada}

\author{C.~Chambers}
\affiliation{TRIUMF, Vancouver, British Columbia, V6T 2A3, Canada}

\author{F.~Maldonado Mill\'{a}n}
\affiliation{TRIUMF, Vancouver, British Columbia, V6T 2A3, Canada}

\author{A.~Mollaebrahimi}
\affiliation{II. Physikalisches Institut, Justus-Liebig-Universität, Gießen, 35392, Germany.}
\affiliation{GSI Helmholtz Center for Heavy Ion Research, Campus Gie\ss en, 35392, Gie\ss en, Germany}

\author{D.~Ray}
\affiliation{TRIUMF, Vancouver, British Columbia, V6T 2A3, Canada}

\author{A.~Weaver}
\affiliation{TRIUMF, Vancouver, British Columbia, V6T 2A3, Canada}

\author{J.~Yu}
\affiliation{GSI Helmholtz Center for Heavy Ion Research, Campus Gie\ss en, 35392, Gie\ss en, Germany}

\author{I.~Dillmann}
\affiliation{TRIUMF, Vancouver, British Columbia, V6T 2A3, Canada}
\affiliation{Department of Physics and Astronomy, University of Victoria, Victoria, British Columbia, V8O 5C2, Canada}

\author{A.~Gezerlis\text{\orcidicon{0000-0003-2232-2484}}}
\email{gezerlis@uoguelph.ca}
\affiliation{Department of Physics, University of Guelph, Guelph, ON N1G 2W1, Canada}

\author{G.~Gwinner}
\affiliation{TRIUMF, Vancouver, British Columbia, V6T 2A3, Canada}
\affiliation{Department of Physics and Astronomy, University of Manitoba, Winnipeg, Manitoba, R3T 2N2 Canada}

\author{A.~O.~Macchiavelli}
\affiliation{Physics Division, Oak Ridge National Laboratory, Oak Ridge, Tennessee 37831, USA}

 \author{S.~Malbrunot-Ettenauer\text{\orcidicon{0009-0004-3323-6500}}}
\email{sette@triumf.ca}
\affiliation{TRIUMF, Vancouver, British Columbia, V6T 2A3, Canada}
\affiliation{Department of Physics, University of Toronto, Toronto, Ontario, M5S 1A7, Canada}

\author{M.P.~Reiter}
\affiliation{School of Physics and Astronomy, The University of Edinburgh, Edinburgh, EH9 3FD, Scotland, UK}

\author{A. A. Kwiatkowski}
\affiliation{TRIUMF, Vancouver, British Columbia, V6T 2A3, Canada}
\affiliation{Department of Physics and Astronomy, University of Victoria, Victoria, British Columbia, V8O 5C2, Canada}

\begin{abstract}

Shell structure in finite quantum systems gives rise to sudden changes in observable properties, while pairing correlations often compete against such discontinuities. The region near the doubly magic nucleus $^{132}$Sn provides a fertile ground for testing the combined effect of shell structure and pairing. Here, we provide a novel phenomenological interpretation of existing mass data in the vicinity of the $Z=50$ and $N=82$ shell closures, which we further investigate by performing original Hartree-Fock-Bogolyubov (HFB) mean-field calculations for even-$Z$ nuclei: we find that the proton shell structure enhances an asymmetry of the neutron odd-even staggering in binding energies. We also report mass measurements of $^{137,138}$Sb, including the first experimental mass determination of $^{138}$Sb, performed using TRIUMF’s Ion Trap for Atomic and Nuclear Science (TITAN). Together with existing experimental data, our results reveal an interplay between shell structure and pairing in odd-$Z$ nuclei which is more challenging to interpret phenomenologically or using HFB, thereby motivating future experimental and theoretical pairing studies in heavy neutron-rich nuclides.
\end{abstract}

\maketitle

\PRLsection{Introduction} Shell structure arises in finite fermionic systems from the Pauli exclusion principle under spatial confinement.
Introduced in the context of distributing electrons across atomic orbitals---a finding that proved paradigmatic for the formulation of quantum mechanics---the concept of shell filling, characterized by pronounced energy gaps, was subsequently found to also govern the structure of atomic nuclei~\cite{mgm:1948, mgm:1949, haxel:1949, brown:2001, caurier:2005, hebeler:2015}, quantum dots~\cite{reimann:2002}, metal nanoclusters~\cite{wilcoxon:2006}, and trapped ultracold Fermi gases~\cite{giorgini:2008}. 
Another hallmark of fermionic systems is the tendency of two fermions to form pairs, typically with zero total angular momentum. The consequences of such pairing are diverse, including the emergence of superconductivity through Cooper pairs~\cite{bennemann:2008} and superfluidity in fermionic systems~\cite{strinati:2008}. 

In low-energy nuclear physics, where protons and neutrons bind to form atomic nuclei, the structure and behavior of these nuclei are governed by both shell effects and nucleon pairing. Shell structure gives rise to a variety of observable phenomena, most notably the enhanced binding associated with so-called magic numbers for protons (\textit{Z}) or neutrons (\textit{N}) reflecting shell closures, as well as systematic patterns in nuclear spins and parities, magnetic moments, excitation energies, and transition strengths between nuclear states. Nuclear pairing, in turn,  is one of the sources behind the well-known odd-even staggering of nuclear binding energies and charge radii in finite nuclei, and also has important implications for the study of superfluidity in neutron stars~\cite{dean:2003}. 

In contrast to other fermionic systems, nuclear pairing energies (typically 2~MeV) are of the same order of magnitude as the energy gaps between nuclear (sub)shells (1–5 MeV), leading to a rich interplay. While shell structure governs the behavior of near-magic nuclei, 
leading to abrupt observable changes across shell closures,
pairing effects are most pronounced in mid-shell regions. In some regions of the nuclear chart, pairing appears to weaken shell effects even near magic numbers, resulting in more gradual changes in nuclear properties. Such competition between shell structure and pairing may become increasingly important in more weakly bound systems, i.e., in nuclei located farther from the valley of stability~\cite{dobaczewski:1996,erler:2012, sorlin:2008}.

In this Letter, we report on the successful mean-field description of an intriguing trend in the odd–even staggering (OES) of binding energies around the doubly magic tin isotope $^{132}$Sn, reported in Ref.~\cite{hakala:2012}, which illustrates this remarkable interplay between nuclear shell structure and pairing far away from stability. We extend our analysis and theoretical modeling to isotopic chains well beyond tin, demonstrating that proton sub-shell closures also shape this interplay. To broaden our study further, we present precision mass measurements of antimony isotopes, consisting of $^{137}$Sb with a 2.5-fold improved precision and a first time determination of $^{138}$Sb, enabling an investigation of the role of unpaired protons in this competition between shell structure and pairing. 

\begin{figure*}[t]
    \centering
    \includegraphics[width=2\columnwidth]{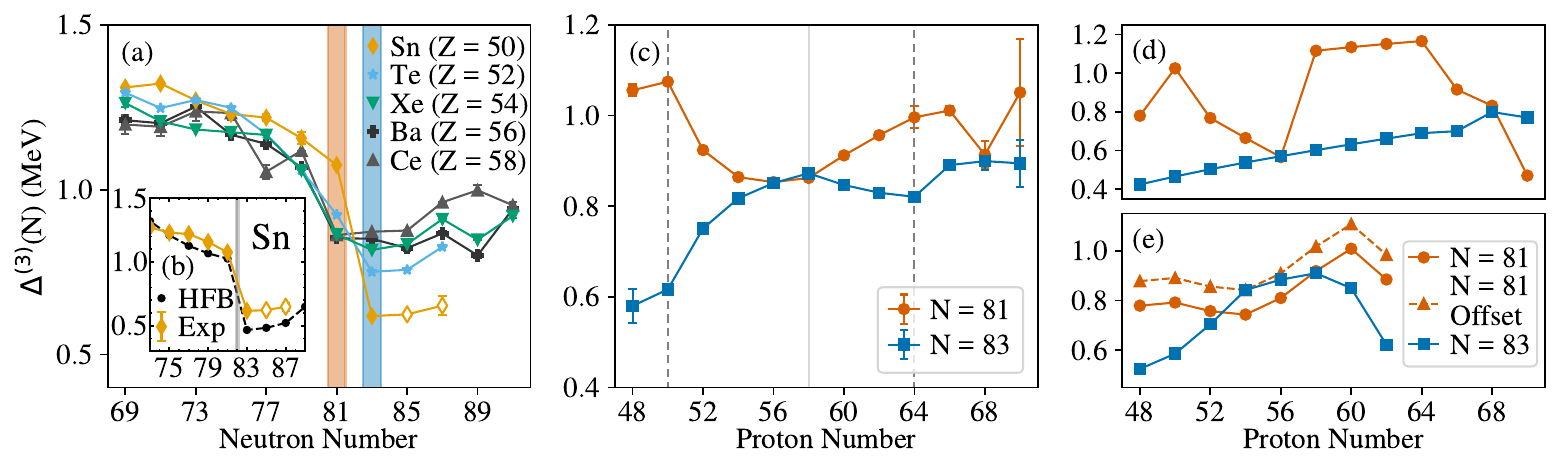}
    \caption{Odd-even staggering of binding energies around $^{132}$Sn. (a)~Experimental $\Delta^{(3)}(N)$ along even-$Z$ isotopic chains for odd-$N$ in the vicinity of the  $N=82$ shell closure. The plotted data is based on Ref.\,\cite{huang2021:ame2020i}, in agreement with recent studies \cite{spataru:2025,kimura:2024,liu:2026}. TITAN measurements involving previously unknown Sn masses \cite{mollaebrahimi:2025}  are indicated as open diamonds. The inset (b) compares experimental $\Delta^{(3)}(N)$ in Sn isotopes with the staggering of the pairing correlation energy $E_{\textrm{corr}}$ from Hartree-Fock-Bogolyubov (HFB) theory. The vertical bands in (a) highlight $\Delta^{(3)}(N=81)$ (orange) and $\Delta^{(3)}(N=83)$ (blue), which are separately displayed in (c) as a function of $Z$, also extending to nuclei not shown in (a), including recently determined Yb masses \cite{beck:2021}. The corresponding theoretical results for spherical and deformed HFB calculations are displayed in (d) and (e), respectively.}
    \label{fig:N_81_83_isotones}
\end{figure*}

\PRLsection{Phenomenological interpretation of existing mass data}
The OES of binding energies, obtained from mass measurements, can be traced to pairing correlations in the nuclear ground-state wavefunction: even-mass nuclei are fully paired and more bound than odd-mass ones \cite{bohr:1958,bohr:book,satula:1998}. The amplitude of the staggering can be extracted by finite difference formulas comparing the binding energy of even- with odd-mass nuclei, typically  via the three-point formula,
\begin{align}
    \Delta^{(3)}(n) = \frac{(-1)^n}{2}\left[B(n+1)-2B(n)+B(n-1)\right]~, \label{eq:oes}
\end{align}
where $n$ is the particle number of one nucleonic species, $N$ or $Z$, and $B(N,Z)$ is the binding energy of the nucleus~\cite{satula:1998,duguet:2001,bender:2003}. In essence, $B(N,Z)$ of even-mass and odd-mass nuclei fall on two separate curves separated by $\Delta^{(3)}$ when plotted as a function of $N$ or $Z$. The OES, i.e., $\Delta^{(3)}$, is then connected to the pairing gap (the energy cost of breaking a Cooper pair). Condensation of Cooper pairs is the mechanism behind fermionic superfluidity, linking $\Delta^{(3)}$ to the nucleus' superfluid properties~\cite{brink:2005,miller:2019,palkanoglou:2020,Duguet:2020}.

Figure\,\ref{fig:N_81_83_isotones}(a) summarizes experimentally known $\Delta^{(3)}(N)$ across isotopic chains of even-$Z$ elements around $^{132}$Sn. As reported in Ref.\,\cite{hakala:2012}, $\Delta^{(3)}(N)$ shows a strong asymmetry between $N\!=\!81$ and $N\!=\!83$ in Sn, but a much weaker one in Te and Xe. We complement the observations of Ref.\,\cite{hakala:2012} by adding data along the Ba and Ce isotopic chains from Ref.\,\cite{huang2021:ame2020i}, as well as from recent mass measurements of even more neutron-rich Sn  isotopes\,\cite{mollaebrahimi:2025}. Neglecting less influential local phenomena, such as the one in Ce at $N=77$ and $N=79$, clear global trends emerge: For smaller $N$, $\Delta^{(3)}(N)$ are found at $\sim1.25$~MeV for all displayed isotopic chains and gradually decrease  with fairly similar slope (at least up to $N\sim77$) when approaching the $N=82$ shell closure. For $N>82$, $\Delta^{(3)}(N)$ remain flat to first order over $N$. Thanks to Ref.~\cite{mollaebrahimi:2025}, this is now also confirmed for Sn, yet the values are strikingly different for each isotopic chain, ranging from $\approx$0.6~MeV in Sn, seemingly increasing with $Z$ until reaching $\approx$0.9~MeV in Ce. Most remarkable, however, is the transition of $\Delta^{(3)}(N)$ across the $N=82$ shell closure, visualized by $\Delta^{(3)}(N=81)$: While Sn exhibits a continuation of the overall downward linear trend within this shell, $\Delta^{(3)}(N=81)$ for Xe, Ba, and Ce is already reduced to the quasi-constant value observed for $\Delta^3(N)$ in the shell above $N=82$. $\Delta^{(3)}(N=81)$ for Te, only two protons above the $Z=50$ shell closure, exhibits an intermediate behavior between Sn and the other displayed isotopic chains.

Reference\,\cite{hakala:2012} examined these relative trends for different $Z$ around the $N=82$ shell closure by comparing $\Delta^{(3)}(N)$ at $N=81$ with those at $N=83$.  Initially adopted for Sn, Te, Xe \cite{hakala:2012}, we follow this perspective and extend it to all 12 even-$Z$ elements for which  $\Delta^{(3)}(N)$ along the $N=81,83$ isotones are known experimentally, see Fig.\,\ref{fig:N_81_83_isotones}(c). As noted in Ref.\,\cite{hakala:2012}, the asymmetry between $\Delta^{(3)}(N=81)$ and $\Delta^{(3)}(N=83)$ is largest for Sn $(Z=50)$ and decreases with increasing $Z$. Interestingly, our extension of this comparison to $Z\geq56$ reveals a non-monotonic behavior: the two $\Delta^{(3)}(N)$ become nearly identical at $Z=56,58$, then diverge again, reaching a maximum difference at $Z=64$, after which this differences decreases once more and vanishes toward $Z=68$. Thus, also considering values for Cd ($Z=48$), the asymmetry between $\Delta^{(3)}(N=81)$ and $\Delta^{(3)}(N=83)$ exhibits local maxima at $Z=50$ and  $Z=64$, see dashed vertical lines in Fig.\,\ref{fig:N_81_83_isotones}(c). According to our single-particle energies obtained via theoretical calculations (see supplemental material), the proton shell gap at $Z=64$ is $\approx 2-3$~MeV, indicative of a pronounced subshell closure at this proton number. In contrast, a near degeneracy ($\lesssim 0.7$~MeV) of the $g_{7/2}$ and $d_{5/2}$ orbitals  suppresses the emergence of a distinct sub-shell closure at $Z=58$, leaving neutron pairing correlations largely unaffected by proton shell effects, see solid gray line in Fig.\,\ref{fig:N_81_83_isotones}(c).

Phenomenologically, one thus concludes that the difference between $\Delta^{(3)}(N=81)$ and $\Delta^{(3)}(N=83)$ is largest at proton (sub-)shell closures, where the underlying shell structure leaves a pronounced imprint on neutron pairing correlations. In the proton mid-shell regions, the $\Delta^{(3)}(N)$ values for the $N=81$ and $N=83$ isotones converge and become nearly identical. This contrasts with the behavior observed for $N<77$, where $\Delta^{(3)}(N)$ remains fairly constant as a function of $Z$, showing neither a pronounced signature of the shell closure $Z=50$ nor a significant reduction in the proton mid-shell region (see Fig.\,3 in Supplemental Material). For example, $\Delta^{(3)}(N=73)$ in Sn, with its closed proton shell, is virtually indistinguishable from that in mid-shell Ce, see also Fig.\,\ref{fig:N_81_83_isotones}(a), while the difference between Sn and Ce is  the most significant feature of both $\Delta^{(3)}(N=81)$ and $\Delta^{(3)}(N=83)$. The latter is traced to two factors that underscore the intricate interplay between shell structure and pairing: (1) an attenuation of $\Delta^{(3)}(N=81)$, consistent with neutron pairing correlations that smear out the neutron shell structure at $N=82$ in the proton-mid shell region, and (2) specifically low  $\Delta^{(3)}(N=83)$ values for $Z=50$ and $Z=64$ at which the proton (sub)shell closures seemingly weaken the (neutron) pairing's impact. 

\PRLsection{Pairing theory}
Following the work in Ref.\,\cite{hakala:2012}, the OES around $^{132}$Sn has attracted attention in shell-model~\cite{brown:2015,coraggio:2013} and Density Functional Theory~\cite{shi:2020} studies. The shell-model calculations identified the importance of the neutron-proton interaction~\cite{coraggio:2013} and the right shell-structure in explaining the relevant behavior. The DFT studies corroborated the importance of the single-particle levels (also see Ref. \cite{orlandi:2018}), also emphasizing that the type of pairing interaction is significant. Crucially, all calculations have focused on the closest neighbors to $^{132}$Sn, that is, Te and Xe isotopes, missing the chance to identify the trend as part of a pattern that repeats for heavier nuclei along the $N=81$ and $83$ isotones. 

In this Letter, we investigate the origin of this pattern by performing original calculations in Hartree-Fock-Bogolyubov (HFB) theory, a mean-field description explicitly including pairing correlations. In HFB, a nuclear ground-state is computed as 
\begin{align}
    E_{\textrm{HFB}}(N,Z)=\textrm{Tr}\left[\epsilon (N,Z) \rho(N,Z) -\frac{1}{2 }\Delta(N,Z)\kappa (N,Z)\right] \label{eq:hfb_energy}
\end{align}

where $\rho(N,Z)$ and $\kappa(N,Z)$ are the normal and anomalous (or pair) density matrices for the nucleus, respectively~\cite{ring:book}. For a given nucleus, the single-particle spectrum $\epsilon_i$ is determined by solving a Schr{\"o}dinger's equation with a spherical or axially deformed Woods-Saxon trapping potential and a spin-orbit term~\cite{gezerlis:2011,palkanoglou:2025a}. The pairing potential is connected to the anomalous density and a two-body interaction as $\Delta_{ij}(N,Z) = \frac{1}{2}\sum_{kl}v_{ijkl}\kappa_{kl}(N,Z)$. A novel aspect of our HFB formulation is that it allows for proton-neutron pairing in singlet and triplet states~\cite{gezerlis:2011,palkanoglou:2025a}. This type of pairing does not arise for most of the neutron-rich isotopes studied here because the proton and neutron Fermi surfaces remain separated by several MeV, making the pairing of particles across them unlikely; yet, as we move towards the proton-rich side of the $N=81$ and $N=83$ isotonic chains, see Fig.\,\ref{fig:N_81_83_isotones}(c), we do find a sizable contribution from proton-neutron (spin-triplet) pairing at the ground states, up to $20\%$ for the most proton-rich isotopes. We describe odd-even isotopes as one-quasiparticle states by imposing an odd number parity to the density matrices~\cite{robledo:2011,gezerlis:2011}, and the odd-odd isotopes as two-quasiparticle states (a marked improvement over the fully paired descriptions of Refs.~\cite{gezerlis:2011,palkanoglou:2025b}).

Using HFB, we calculate the pairing correlation energy (the pairing's contribution to the binding energy), $E_{\textrm{corr}}(N,Z)=E_{\textrm{HFB}}(N,Z)-E_{\textrm{NS}}(N,Z)$, where $E_{\textrm{NS}}$ is the energy of the normal (non-superfluid) state, obtained from Eq.\,\ref{eq:hfb_energy} for vanishing $\kappa$. Next, we determine the OES based on $E_{\textrm{corr}}$: since any slowly changing part of the binding energy approximately drops out in Eq.\,\ref{eq:oes}, this approach captures the pairing's contribution to the OES. We validate the calculated OES values against experimental data for Sn isotopes, see Fig.\,\ref{fig:N_81_83_isotones}(b). Excellent agreement is obtained, despite the absence of any \textit{ad hoc} fitting parameters in our HFB model.

The behavior of $\Delta^{(3)}(N)$ seen in Fig.\,\ref{fig:N_81_83_isotones}(a) can thus be interpreted as arising from the growth of the neutron pairing correlations when the proton number increases (see supplemental material), reflecting the importance of interactions across nucleonic species. The neutron pairing correlations, in turn, smear the neutron shell closure at $N=82$, thereby leading to a less-pronounced asymmetry of the pairing gaps across it. This picture is laid out in Fig.\,\ref{fig:N_81_83_isotones}(d), which shows the OES for the $N=81$ and $N=83$ isotonic chains (on either side of the shell closure), using spherical HFB. Theory results for $\Delta^{(3)}(N)$ qualitatively exhibit the same features at (sub)shell closures as the experimental data in Fig.\,\ref{fig:N_81_83_isotones}(c), further corroborating our interpretation. Spherical HFB  reproduces particularly well the trends of experimental $\Delta^{(3)}(N=81)$ starting form  $Z=48$ up to $Z=56$, where $\Delta^{(3)}(N=81)\approx \Delta^{(3)}(N=83)$. For $\Delta^{(3)}(N=83)$, theory likewise captures the overall increase with $Z$, but predict a nearly linear dependence that differs markedly from experiment. The latter deficiency is overcome by the deformed HFB calculation (see supplemental material), which accurately describes the curvature and absolute values of $\Delta^{(3)}(N=83)$ until $Z=60$, see Fig.\,\ref{fig:N_81_83_isotones}(e). Within the same range of $Z$, the relative trends in $\Delta^{(3)}(N=81)$ are qualitatively reproduced, too, though the absolute value at $Z=54$ is underestimated by approximately 0.1~MeV, see (red) dashed curve in Fig.\,\ref{fig:N_81_83_isotones}(e).

Having demonstrated that our novel theoretical HFB calculations reproduce well the available experimental data, we seek more demanding tests of their predictive power. Extending the comparison of Fig.\,\ref{fig:N_81_83_isotones}(c) towards lower (higher) $Z$ is, however, experimentally difficult, as the corresponding nuclei lie increasingly far from stability. Through mass measurements of Sb isotopes, we thus turn to odd-$Z$ nuclei, where the presence of an unpaired proton provides a more challenging setting for theory.

\begin{figure}[t]
    \centering
    \includegraphics[width=1\columnwidth]{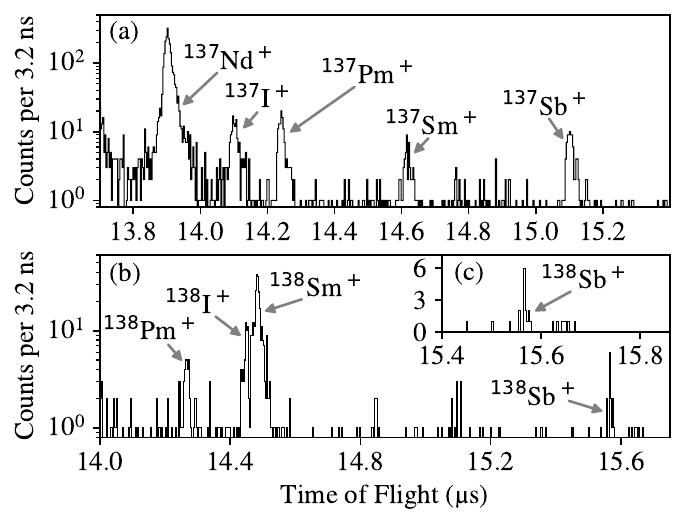}
    \caption{Time-of-Flight spectra (TOF) recorded by TITAN during the Sb measurement campaign, shown as a function of ion counts versus time elapsed since ion extraction from the MR-TOF-MS analyzer, for (a) $A=137$ with 678 turns in the analyzer 
    and (b) $A=138$ with 703 turns providing a mass resolving power of $R = m/\delta m\approx$390,000. Inset (c) shows a zoom around the peak associated with $^{138}$Sb$^+$ ions. Peaks corresponding to the main calibrants $^{137}$Cs$^+$ and $^{138}$Ba$^+$, respectively, are outside of the displayed TOF range.}
    \label{fig:137138Sb}
\end{figure} 

\PRLsection{Experiment}
High-precision mass measurements of Sb isotopes in the vicinity of $^{132}$Sn are performed using TRIUMF's Ion Trap for Atomic and Nuclear Science (TITAN) \cite{kwiatkowski:2024} located within the Isotope Separator and Accelerator (ISAC) \cite{ball:2016, ball:2020} facility at TRIUMF. Short-lived radionuclei are produced by impinging a 480\,MeV, $\approx$18\,$\mu$A proton beam onto a uranium carbide target. Neutral atoms diffusing out of the production target are ionized using either TRIUMF’s Resonant Laser Ion Source (TRILIS) \cite{lassen:2023} in the case of Sb, or surface ionization \cite{dombsky:2000} for all other species delivered to TITAN. Following electrostatic acceleration, the resulting radioactive ion beam is filtered by a magnetic dipole mass separator \cite{bricault:2002} before being injected into TITAN's radiofrequency quadrupole cooler-buncher (RFQcb) \cite{brunner:2012}. There, the ions are cooled, accumulated, and subsequently transferred as an ion bunch to a multiple-reflection time-of-flight mass spectrometer (MR-TOF-MS) \cite{jesch:2015}, where the mass measurements are carried out. The latter follow a well-established procedure \cite{leistenschneider:2018, reiter:2021} and incorporate the use of mass-selective retrapping \cite{dickel:2017} to further clean the beam of isobaric contaminants at each atomic mass unit $A$. The resulting TOF spectra for $A=137$ and $A=138$ are shown in Fig.\,\ref{fig:137138Sb}, including TOF peaks associated with $^{137}$Sb$^+$ (103 counts) and  $^{138}$Sb$^+$ ions (14 counts). 

Once the data acquisition is complete, the masses of the ion species are determined through analysis with our \textsc{emgfit} package \cite{paul:2020} which fits hyper-EMG functions to the TOF data \cite{purushothaman:2017}. Statistical and systematic uncertainties are estimated following previously established procedures outlined in Ref.\,\cite{andres:2019, paul:2021}. Due to the low production yields of the ions of interest, the statistical uncertainty of $\approx$20~keV for $^{137}$Sb and $\approx$68~keV for $^{138}$Sb, respectively, dominate over the systematic error which is assigned as $\delta m/m$ = 1 $\times$ 10$^{-7}$. As an accuracy check, all labeled peaks in Fig.\,\ref{fig:137138Sb} are compared with literature values reported in Ref.\,\cite{huang2021:ame2020i}, showing good agreement.
\begin{table}[!t]
\caption{Atomic masses of Sb isotopes as determined by TITAN. Data are listed as an atomic mass ratio ($m_{\text{ion}}/m_{\text{calibrant}}$) to the main TOF-calibrating  species and as the mass excess (ME); numbers in parentheses represent uncertainties. For comparison, the Atomic Mass Evaluation reports the ME of $^{137}$Sb as -60060(50)~keV/$c^2$ \cite{huang2021:ame2020i}. The mass of $^{138}$Sb is measured for the first time.}
\label{tab:Mass}
\centering
\begin{tabular}{c c c c}
\hline \hline \\ [-1em]
Species & Calibrant & Mass Ratio & ME$_{\textrm{TITAN}}$ (keV/c$^{2}$) \\
[-1em]\\
\hline \\ [-1em]
$^{137}$Sb & $^{137}$Cs  & 1.000 207 858 (156) & -60037 (20) \\
[-1em]\\
$^{138}$Sb & $^{138}$Ba & 1.000 263 755 (525) & -54379 (68) \\
[-1em]\\
\hline \hline
\end{tabular}
\end{table}
Table\,\ref{tab:Mass} summarizes the experimental Sb results. Our final atomic mass for $^{137}$Sb agrees with the literature value, while reducing the uncertainty by a factor of approximately 2.5. The mass of $^{138}$Sb was previously unknown and is determined here for the first time.  These results reduce the uncertainty of $\Delta^{(3)}(N=85)$ in Sb and allow for the first experimental evaluation of Sb's $\Delta^{(3)}(N=86)$.

\begin{figure}[t]
    \centering
    \includegraphics[width=1\columnwidth]{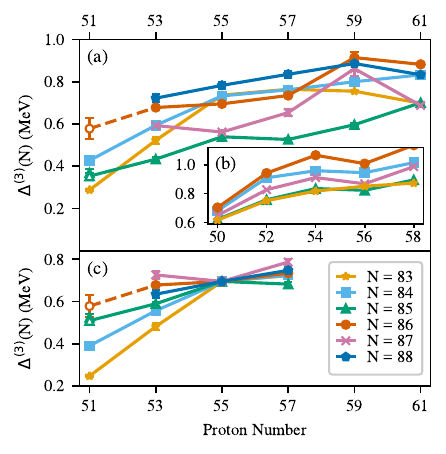}
    \caption{Experimental odd-even staggering of binding energies $\Delta^{(3)}(N)$ above $^{132}$Sn for isotonic chains $N=83$ to $N=88$, for (a) odd-$Z$ and (b) even-$Z$ nuclei. Data points improved/introduced in the present work are indicated by open symbols. In (c), data shown in (a) are vertically shifted by isotonic-specific offsets such that all chains are aligned at $Z=55$ to the value $\Delta^{(3)}(N=86)$. See text for details.}
    \label{fig:Odd_N_81_83_isotones}
\end{figure}

\PRLsection{OES including odd-$Z$ nuclei}  
Figure\,\ref{fig:Odd_N_81_83_isotones}(a) depicts experimental $\Delta^{(3)}(N)$ for nuclei with odd proton numbers ($Z > 50$) along the isotonic chains $N = 83$–$88$. We first observe that, for $Z \ge 55$, all isotonic $\Delta^{(3)}(N)$ curves increase only gradually with $Z$, resulting in a relatively flat behavior (with the exception of a kink at $Z=59$ for $N=86$ and $N=87$, indicative of a local, unrelated effect). This is consistent with the trends seen in even-$Z$ isotonic chains for $Z\ge54$, see Fig.\,\ref{fig:Odd_N_81_83_isotones}(b). For $Z \leq 53$, almost all $\Delta^{(3)}(N)$ values are significantly reduced, with the effect becoming progressively stronger toward the proton shell closure at $Z=50$.

As revealed by our new Sb mass data, this weakening of $\Delta^{(3)}(N)$ is not nearly as strong for odd-$Z$ $\Delta^{(3)}(N=85)$ and $\Delta^{(3)}(N=86)$. For illustrative purposes, the isotonic $\Delta^{(3)}(N)$ curves shown in Fig.\,\ref{fig:Odd_N_81_83_isotones}(c) are vertically shifted such that their respective value at $Z = 55$ aligns with the $N = 86$ curve. We observe that the slope in $\Delta^{(3)}(N)$ between $Z=51$ and $Z=55$ greatly varies for different isotonic chains: For $N=83$, $\Delta^{(3)}(N)$ increases steeply with $Z$. With each added neutron, however, this increase becomes progressively weaker until the corresponding curves are nearly flat by $N=85$ and $N=86$, consistent with their respective trend above $Z\ge55$.

Phenomenologically, the differences in $\Delta^{(3)}(N)$ between the various $N$ values at $Z=51$ and $Z=53$ could, at first sight, be understood as a consequence of $N=83$ corresponding to a single valence neutron outside the closed neutron shell, where pairing correlations are expected to be strongly suppressed; the addition of further neutrons would progressively enhance pairing correlations and thus increase $\Delta^{(3)}(N)$, as indeed observed for odd-$Z$ nuclei with $Z=53$ and - given our new Sb data - even more pronounced in $Z=51$.  This picture is, however, incomplete: $\Delta^{(3)}(N)$ in even-$Z$ nuclei remain low for all $Z=50$, independently of the number of valence neutrons above $N=82$, see again Fig.\,\ref{fig:Odd_N_81_83_isotones}(b). Additionally, this argument would not explain why the isotonic chains exhibit a rapid increase in $\Delta^{(3)}(N)$ with proton number before reaching saturation at $Z=54$ or $Z=55$, respectively. As discussed in connection with Fig.\,\ref{fig:N_81_83_isotones}(c), the reduced $\Delta^{(3)}(N)$ values between $Z=50$ and $Z=53$ may instead reflect a more general impact of the $Z=50$ proton shell closure on neutron pairing. Intriguingly, as our new data indicates, such a shell-structure effect appears to be overcome by stronger neutron pairing in more neutron-rich Sb ($Z=51$) and I ($Z=53$) isotopes, in which $\Delta^{(3)}(N=85)$ and $\Delta^{(3)}(N=86)$ are as large as their associated isotonic values for $Z\geq55$. 

As already noted, our deformed HFB calculations reproduce well the increase of even-$Z$ $\Delta^{(3)}(N=83)$ from initially low values at the proton shell closure $Z=50$ to its saturation in the mid-shell region around $Z=58$. Similar agreement between HFB results and experiment in this $Z$ range is also found for $N>83$. Within our model, the increase in $\Delta^{(3)}(N)$ with Z - beginning from their reduced values at the proton shell closure - arises from an enhancement in the neutron-neutron pairing correlations induced by interactions across nucleonic species. For nuclei with odd $Z$, however, both spherical and deformed HFB calculations yield results inconsistent with experimental $\Delta^{(3)}(N)$. Various attempts to improve our HFB model did not lead to a satisfactory description, motivating more comprehensive extensions of our framework to odd-odd nuclei in the region.

\PRLsection{Summary}
We have investigated the odd-even staggering of nuclear binding energies in the vicinity of the $Z=50$ and $N=82$ shell closures. We find evidence that the proton shell structure influences neutron pairing at both $Z=50$ and $Z=64$, enhancing an asymmetry of the neutron OES across $N=82$. This effect is partly driven by unusually low OES values at or near proton shell closures for $N\geq83$. Our new mass measurements of $^{137,138}$Sb reveal that, for $Z=51$ these values increase with the addition of neutrons, while a similar recovery toward larger OES is generally observed in the proton mid-shell region, where the $N=82$ signature in the OES is washed out. HFB calculations successfully reproduce the latter phenomena, highlighting the importance of interactions across nucleonic species. The former behavior is more challenging to interpret phenomenologically and within HFB theory, motivating further theoretical and experimental efforts, notably mass measurements of more neutron-rich isotopes such as of Sn, Sb and I, to achieve a comprehensive understanding of this interplay between shell structure and pairing.

\PRLsection{Acknowledgments}
We would like to thank and TRIUMF's technical teams, in particular Targets and Ion Source group and Beam Delivery, the TITAN Collaboration, as well as M. Good for his technical assistance, and P.-G. Reinhard for insightful conversations. TITAN is funded by the Natural Sciences and Engineering Research Council (NSERC) of Canada and through TRIUMF by the National Research Council (NRC) of Canada. R.~S. acknowledges support from the NSERC Graduate Research Scholarship. The work of A.G. was supported by NSERC and the Canada Foundation for Innovation (CFI). Computational resources have been provided by Compute Ontario through the Digital Research Alliance of Canada, and by the National Energy Research Scientific Computing Center (NERSC), which is supported by the U.S. Department of Energy, Office of Science, under contract No. DE-AC02-05CH11231. M.P.R. was supported by UKRI STFC under grant agreement no.\ ST/Y000293/1.

\bibliography{bibliography}

@article{huang2021:ame2020i,
  title   = {The {AME} 2020 atomic mass evaluation ({I}). {Evaluation} of input data, and adjustment procedures},
  author  = {Huang, W. J. and Wang, Meng and Kondev, F. G. and Audi, G. and Naimi, S.},
  journal = {Chinese Physics C},
  volume  = {45},
  number  = {3},
  pages   = {030002},
  year    = {2021},
  doi     = {10.1088/1674-1137/abddaf}
}

@article{palkanoglou:2025a,
  title={Spin-Triplet Pairing in Heavy Nuclei Is Stable against Deformation},
  author={Palkanoglou, Georgios and Stuck, Michael and Gezerlis, Alexandros},
  journal={Physical Review Letters},
  volume={134},
  number={3},
  pages={032501},
  year={2025},
  publisher={APS}
}

@article{palkanoglou:2025b,
  title={Symmetry properties of pair correlations in heavy deformed nuclei},
  author={Palkanoglou, Georgios and Gezerlis, Alexandros},
  journal={arXiv preprint arXiv:2505.08879},
  year={2025}
}

@article{palkanoglou:2020,
  title={From odd-even staggering to the pairing gap in neutron matter},
  author={Palkanoglou, Georgios and Diakonos, Fotis K and Gezerlis, Alexandros},
  journal={Physical Review C},
  volume={102},
  number={6},
  pages={064324},
  year={2020},
  publisher={APS}
}

@article{gezerlis:2011,
  title={Mixed-spin pairing condensates in heavy nuclei},
  author={Gezerlis, Alexandros and Bertsch, GF and Luo, YL},
  journal={Physical review letters},
  volume={106},
  number={25},
  pages={252502},
  year={2011},
  publisher={APS}
}

@article{ball:2016,
year = {2016},
month = {Jul},
publisher = {IOP Publishing},
volume = {91},
number = {9},
pages = {093002},
author = {G. C. Ball and G. Hackman and R. Krücken},
title = {The {TRIUMF-ISAC} facility: two decades of discovery with rare isotope beams},
journal = {Physica Scripta}
}

@article{ball:2020,
author = {Ball, G. C. and Dillmann, I. and Garnsworthy, A. and Gwinner, G. and Kanungo, R. and Morris, G. and Ruiz, C.},
title = {The {TRIUMF}-{ISAC} Facility: Recent Highlights in {RIB} Science and Future Prospects with {ARIEL}},
journal = {Nuclear Physics News},
volume = {30},
number = {4},
pages = {27--32},
year = {2020},
publisher = {Taylor \& Francis}
}

@article{kwiatkowski:2024,
author = {Kwiatkowski, A. A. and Dilling, J. and Malbrunot-Ettenauer, S. and others},
title = {15 years of precision mass measurements at {TITAN}},
Journal = {The European Physics Journal A},
volume = {60},
number = {87},
year = {2024}
}

@article{lassen:2023,
title = {Developments at TRIUMF’s laser resonance ionization ion source \& multi-element operation},
journal = {Nuclear Instruments and Methods in Physics Research Section B: Beam Interactions with Materials and Atoms},
volume = {541},
pages = {137-140},
year = {2023},
issn = {0168-583X},
author = {J. Lassen and R. Li and M. Mostamand and A. Gacsbaranyi and P. Kunz and C. Babcock and D. Bishop and A. Teigelhöfer and F. Ames and A. Gottberg},
}

@article{dombsky:2000,
    author = {Dombsky, M. and Bishop, D. and Bricault, P. and Dale, D. and Hurst, A. and Jayamanna, K. and Keitel, R. and Olivo, M. and Schmor, P. and Stanford, G.},
    title = {{Commissioning and initial operation of a radioactive beam ion source at ISAC}},
    journal = {Review of Scientific Instruments},
    volume = {71},
    number = {2},
    pages = {978-980},
    year = {2000},
}

@article{brunner:2012,
  title={{TITAN's digital RFQ ion beam cooler and buncher, operation and performance}},
  author={Brunner, T. and Smith, M. J. and Brodeur, M. and Ettenauer, S. and Gallant, A. T. and Simon, V. V. and Chaudhuri, A. and Lapierre, A. and Man{\'e}, E. and Ringle, R. and others},
  journal={Nuclear Instruments and Methods in Physics Research Section A: Accelerators, Spectrometers, Detectors and Associated Equipment},
  volume={676},
  pages={32--43},
  year={2012},
  publisher={Elsevier}
}

@article{reiter:2021,
  title={Commissioning and performance of {TITAN}’s {Multiple-Reflection Time-of-Flight Mass-Spectrometer} and isobar separator},
  author={Reiter, M. P. and San Andr{\'e}s, S. A. and Bergmann, J. and Dickel, T. and Dilling, J. and Jacobs, A. and Kwiatkowski, A. A. and Pla{\ss}, W. R. and Scheidenberger, C. and Short, D. and others},
  journal={Nuclear Instruments and Methods in Physics Research Section A: Accelerators, Spectrometers, Detectors and Associated Equipment},
  volume={1018},
  pages={165823},
  year={2021},
  publisher={Elsevier}
}

@article{leistenschneider:2018,
  title = {Dawning of the {$N=32$} Shell Closure Seen through Precision Mass Measurements of Neutron-Rich Titanium Isotopes},
  author = {Leistenschneider, E. and Reiter, M. P. and Ayet San Andr\'es, S. and Kootte, B. and Holt, J. D. and Navr\'atil, P. and Babcock, C. and Barbieri, C. and Barquest, B. R. and Bergmann, J. and Bollig, J. and Brunner, T. and Dunling, E. and Finlay, A. and Geissel, H. and Graham, L. and Greiner, F. and Hergert, H. and Hornung, C. and Jesch, C. and Klawitter, R. and Lan, Y. and Lascar, D. and Leach, K. G. and Lippert, W. and McKay, J. E. and Paul, S. F. and Schwenk, A. and Short, D. and Simonis, J. and Som\`a, V. and Steinbr\"ugge, R. and Stroberg, S. R. and Thompson, R. and Wieser, M. E. and Will, C. and Yavor, M. and Andreoiu, C. and Dickel, T. and Dillmann, I. and Gwinner, G. and Pla\ss{}, W. R. and Scheidenberger, C. and Kwiatkowski, A. A. and Dilling, J.},
  journal = {Physical Review Letters},
  volume = {120},
  issue = {6},
  pages = {062503},
  numpages = {7},
  year = {2018},
  publisher = {American Physical Society},
}

@article{dickel:2017,
author = {Dickel, Timo and Plaß, Wolfgang R. and Lippert, Wayne and Lang, Johannes and Yavor, Mikhail I. and Geissel, Hans and Scheidenberger, Christoph},
title = {Isobar Separation in a Multiple-Reflection Time-of-Flight Mass Spectrometer by Mass-Selective Re-Trapping},
journal = {Journal of the American Society for Mass Spectrometry},
volume = {28},
number = {6},
pages = {1079-1090},
year = {2017}
}

@software{paul:2020,
  author = {Paul, Stefan F.},
  title = {emgfit - Fitting of time-of-flight mass spectra with hyper-EMG models},
  year         = 2020,
  publisher    = {Zenodo},
  version      = {v0.3.5},
  url = {https://zenodo.org/records/10552901}
}

@article{andres:2019,
  title = {High-resolution, accurate multiple-reflection time-of-flight mass spectrometry for short-lived, exotic nuclei of a few events in their ground and low-lying isomeric states},
  author = {Ayet San Andr\'es, Samuel and Hornung, Christine and Ebert, Jens and Pla\ss{}, Wolfgang R. and Dickel, Timo and Geissel, Hans and Scheidenberger, Christoph and Bergmann, Julian and Greiner, Florian and Haettner, Emma and Jesch, Christian and Lippert, Wayne and Mardor, Israel and Miskun, Ivan and Patyk, Zygmunt and Pietri, Stephane and Pihktelev, Alexander and Purushothaman, Sivaji and Reiter, Moritz P. and Rink, Ann-Kathrin and Weick, Helmut and Yavor, Mikhail I. and Bagchi, Soumya and Charviakova, Volha and Constantin, Paul and Diwisch, Marcel and Finlay, Andrew and Kaur, Satbir and Kn\"obel, Ronja and Lang, Johannes and Mei, Bo and Moore, Iain D. and Otto, Jan-Hendrik and Pohjalainen, Ilkka and Prochazka, Andrej and Rappold, Christophe and Takechi, Maya and Tanaka, Yoshiki K. and Winfield, John S. and Xu, Xiaodong},
  journal = {Physical Review C},
  volume = {99},
  issue = {6},
  pages = {064313},
  numpages = {21},
  year = {2019},
  publisher = {American Physical Society},
}

@article{paul:2021,
  title = {{Mass measurements of $^{60\text{--}63}\mathrm{Ga}$ reduce x-ray burst model uncertainties and extend the evaluated $T=1$ isobaric multiplet mass equation}},
  author = {Paul, S. F. and Bergmann, J. and Cardona, J. D. and Dietrich, K. A. and Dunling, E. and Hockenbery, Z. and Hornung, C. and Izzo, C. and Jacobs, A. and Javaji, A. and Kootte, B. and Lan, Y. and Leistenschneider, E. and Lykiardopoulou, E. M. and Mukul, I. and Murb\"ock, T. and Porter, W. S. and Silwal, R. and Smith, M. B. and Ringuette, J. and Brunner, T. and Dickel, T. and Dillmann, I. and Gwinner, G. and MacCormick, M. and Reiter, M. P. and Schatz, H. and Smirnova, N. A. and Dilling, J. and Kwiatkowski, A. A.},
  journal = {Physical Review C},
  volume = {104},
  issue = {6},
  pages = {065803},
  numpages = {14},
  year = {2021},
  month = {Dec},
  publisher = {American Physical Society},
}

@article{bricault:2002,
title = {{TRIUMF-ISAC target station and mass separator commissioning}},
journal = {Nuclear Physics A},
volume = {701},
number = {1},
pages = {49-53},
year = {2002},
note = {5th International Conference on Radioactive Nuclear Beams},
issn = {0375-9474},
author = {Pierre Bricault and Richard Baartman and Marik Dombsky and Andrew Hurst and Clive Mark and Guy Stanford and Paul Schmor}
}

@article{purushothaman:2017,
title = {Hyper-EMG: A new probability distribution function composed of Exponentially Modified Gaussian distributions to analyze asymmetric peak shapes in high-resolution time-of-flight mass spectrometry},
journal = {International Journal of Mass Spectrometry},
volume = {421},
pages = {245-254},
year = {2017},
issn = {1387-3806},
author = {S. Purushothaman and S. {Ayet San Andrés} and J. Bergmann and T. Dickel and J. Ebert and H. Geissel and C. Hornung and W.R. Plaß and C. Rappold and C. Scheidenberger and Y.K. Tanaka and M.I. Yavor}
}

@article{mollaebrahimi:2025,
  title = {Precision Mass Measurements Reveal Low Neutron Pairing in Tin beyond ${N}=82$ and Its Impact on Stellar Nucleosynthesis},
  author = {Mollaebrahimi, A. and Walls, C. and Dickel, T. and Miyagi, T. and Sieverding, A. and Andreoiu, C. and Ash, J. and Ashrafkhani, B. and Belosevic, I. and Bergman, J. and Brown, C. and Brunner, T. and Cardona, J. and Egoriti, L. and Gelinas, G. and Gwinner, G. and Hockenbery, Z. and Holt, J. D. and Jacobs, A. and Kakkar, S. and Kootte, B. and Lassen, J. and Lykiardopoulou, E. M. and Malbrunot-Ettenauer, S. and Mart\'{\i}nez-Pinedo, G. and Paul, S. F. and Pla\ss{}, W. R. and Reiter, M. P. and Ridley, A. and Scheidenberger, C. and Thompson, R. I. and Wang, Y. and Wieser, M. E. and Kwiatkowski, A. A.},
  journal = {Physical Review Letters},
  volume = {134},
  issue = {23},
  pages = {232701},
  numpages = {9},
  year = {2025},
  month = {Jun},
  publisher = {American Physical Society}
}

@article{hakala:2012,
  title = {Precision Mass Measurements beyond $^{132}\mathrm{Sn}$: Anomalous Behavior of Odd-Even Staggering of Binding Energies},
  author = {Hakala, J. and Dobaczewski, J. and Gorelov, D. and Eronen, T. and Jokinen, A. and Kankainen, A. and Kolhinen, V. S. and Kortelainen, M. and Moore, I. D. and Penttil\"a, H. and Rinta-Antila, S. and Rissanen, J. and Saastamoinen, A. and Sonnenschein, V. and \"Ayst\"o, J.},
  journal = {Phys. Rev. Lett.},
  volume = {109},
  issue = {3},
  pages = {032501},
  numpages = {5},
  year = {2012},
  month = {Jul},
  publisher = {American Physical Society},
  doi = {10.1103/PhysRevLett.109.032501},
  url = {https://link.aps.org/doi/10.1103/PhysRevLett.109.032501}
}

@article{caurier:2005,
  title = {The shell model as a unified view of nuclear structure},
  author = {Caurier, E. and Mart\'{\i}nez-Pinedo, G. and Nowacki, F. and Poves, A. and Zuker, A. P.},
  journal = {Rev. Mod. Phys.},
  volume = {77},
  issue = {2},
  pages = {427--488},
  numpages = {0},
  year = {2005},
  month = {Jun},
  publisher = {American Physical Society},
  doi = {10.1103/RevModPhys.77.427},
  url = {https://link.aps.org/doi/10.1103/RevModPhys.77.427}
}

@article{hebeler:2015,
   author = "Hebeler, K. and Holt, J.D. and Menéndez, J. and Schwenk, A.",
   title = "Nuclear Forces and Their Impact on Neutron-Rich Nuclei and Neutron-Rich Matter", 
   journal= "Ann. Rev. Nucl. Part. Sci.",
   year = "2015",
   volume = "65",
   number = "Volume 65, 2015",
   pages = "457-484",
   doi = "https://doi.org/10.1146/annurev-nucl-102313-025446",
   url = "https://www.annualreviews.org/content/journals/10.1146/annurev-nucl-102313-025446",
   publisher = "Annual Reviews",
   issn = "1545-4134",
   type = "Journal Article",
  }

@Article{wilcoxon:2006,
author ="Wilcoxon, J. P. and Abrams, B. L.",
title  ="Synthesis{,} structure and properties of metal nanoclusters",
journal  ="Chem. Soc. Rev.",
year  ="2006",
volume  ="35",
issue  ="11",
pages  ="1162-1194",
publisher  ="The Royal Society of Chemistry",
doi  ="10.1039/B517312B",
url  ="http://dx.doi.org/10.1039/B517312B"}

@article{giorgini:2008,
  title = {Theory of ultracold atomic Fermi gases},
  author = {Giorgini, Stefano and Pitaevskii, Lev P. and Stringari, Sandro},
  journal = {Rev. Mod. Phys.},
  volume = {80},
  issue = {4},
  pages = {1215--1274},
  numpages = {0},
  year = {2008},
  month = {Oct},
  publisher = {American Physical Society},
  doi = {10.1103/RevModPhys.80.1215},
  url = {https://link.aps.org/doi/10.1103/RevModPhys.80.1215}
}

@article{reimann:2002,
  title = {Electronic structure of quantum dots},
  author = {Reimann, Stephanie M. and Manninen, Matti},
  journal = {Rev. Mod. Phys.},
  volume = {74},
  issue = {4},
  pages = {1283--1342},
  numpages = {0},
  year = {2002},
  month = {Nov},
  publisher = {American Physical Society},
  doi = {10.1103/RevModPhys.74.1283},
  url = {https://link.aps.org/doi/10.1103/RevModPhys.74.1283}
}

@article{strinati:2008,
title = {The {BCS–BEC} crossover: From ultra-cold Fermi gases to nuclear systems},
journal = {Phys. Rept.},
volume = {738},
pages = {1-76},
year = {2018},
issn = {0370-1573},
doi = {https://doi.org/10.1016/j.physrep.2018.02.004},
url = {https://www.sciencedirect.com/science/article/pii/S0370157318300267},
author = {Giancarlo Calvanese Strinati and Pierbiagio Pieri and Gerd Röpke and Peter Schuck and Michael Urban}
}

@book {bennemann:2008,
     TITLE = {Superconductivity},
     EDITOR = {Bennemann, K. H. and Ketterson, J. B.},
     PUBLISHER = {Springer-Verlag },
      YEAR = {2008},
}

@article{dean:2003,
  title = {Pairing in nuclear systems: from neutron stars to finite nuclei},
  author = {Dean, D. J. and Hjorth-Jensen, M.},
  journal = {Rev. Mod. Phys.},
  volume = {75},
  issue = {2},
  pages = {607--656},
  numpages = {0},
  year = {2003},
  month = {Apr},
  publisher = {American Physical Society},
  doi = {10.1103/RevModPhys.75.607},
  url = {https://link.aps.org/doi/10.1103/RevModPhys.75.607}
}

@article{moller:2016,
  title={Nuclear ground-state masses and deformations: FRDM (2012)},
  author={M{\"o}ller, P and Sierk, Arnold John and Ichikawa, Takatoshi and Sagawa, Hiroyuki},
  journal={Atomic Data and Nuclear Data Tables},
  volume={109},
  pages={1--204},
  year={2016},
  publisher={Elsevier}
}

@inproceedings{brown:2015,
  title={The Nuclear Pairing Gap--How Low Can It Go?},
  author={Brown, B Alex},
  booktitle={Journal of Physics: Conference Series},
  volume={580},
  number={1},
  pages={012016},
  year={2015}
}

@article{coraggio:2013,
  title={{Behavior of odd-even mass staggering around $^{132}$Sn}},
  author={Coraggio, L and Covello, A and Gargano, A and Itaco, Nunzio},
  journal={Physical Review C},
  volume={88},
  number={4},
  pages={041304},
  year={2013},
  publisher={APS}
}

@article{shi:2020,
  title={{Abnormal odd-even staggering behavior around $^{132}$Sn studied by density functional theory}},
  author={Shi, Haoqiang and Wang, Xiao-Bao and Dong, Guo-Xiang and Wang, Hualei},
  journal={Chinese Physics C},
  volume={44},
  number={9},
  pages={094108},
  year={2020},
  publisher={Chinese Physical Society and the Institute of High Energy Physics of the~…}
}

@article{duguet:2020,
  title={Zero-pairing limit of Hartree-Fock-Bogoliubov reference states},
  author={Duguet, Thomas and Bally, Benajamin and Tichai, A},
  journal={Physical Review C},
  volume={102},
  number={5},
  pages={054320},
  year={2020},
  publisher={APS}
}

@article{robledo:2011,
  title={{Application of the gradient method to Hartree-Fock-Bogoliubov theory}},
  author={Robledo, LM and Bertsch, GF},
  journal={Physical Review C},
  volume={84},
  number={1},
  pages={014312},
  year={2011},
  publisher={APS}
}

@article{duguet:2001,
  title={Pairing correlations. II. Microscopic analysis of odd-even mass staggering in nuclei},
  author={Duguet, Thomas and Bonche, Paul and Heenen, P-H and Meyer, Jacques},
  journal={Physical Review C},
  volume={65},
  number={1},
  pages={014311},
  year={2001},
  publisher={APS}
}

@article{bender:2003,
  title={Self-consistent mean-field models for nuclear structure},
  author={Bender, Michael and Heenen, Paul-Henri and Reinhard, Paul-Gerhard},
  journal={Reviews of Modern Physics},
  volume={75},
  number={1},
  pages={121},
  year={2003},
  publisher={APS}
}

@article{satula:1998,
  title={Odd-even staggering of nuclear masses: Pairing or shape effect?},
  author={Satu{\l}a, W and Dobaczewski, J and Nazarewicz, W},
  journal={Physical Review Letters},
  volume={81},
  number={17},
  pages={3599},
  year={1998},
  publisher={APS}
}

@book{brink:2005,
  title={Nuclear superfluidity: pairing in finite systems},
  author={Brink, David M and Broglia, Ricardo A},
  year={2005},
  publisher={Cambridge University Press}
}

@article{miller:2019,
  title={Proton superfluidity and charge radii in proton-rich calcium isotopes},
  author={Miller, Andrew J and Minamisono, Kei and Klose, A and Garand, David and Kujawa, C and Lantis, JD and Liu, Yuan and Maa{\ss}, B and Mantica, PF and Nazarewicz, W and others},
  journal={Nature physics},
  volume={15},
  number={5},
  pages={432--436},
  year={2019},
  publisher={Nature Publishing Group UK London}
}

@article{erler:2012,
  title={The limits of the nuclear landscape},
  author={Erler, Jochen and Birge, Noah and Kortelainen, Markus and Nazarewicz, Witold and Olsen, Erik and Perhac, Alexander M and Stoitsov, Mario},
  journal={Nature},
  volume={486},
  number={7404},
  pages={509--512},
  year={2012},
  publisher={Nature Publishing Group UK London}
}

@article{dobaczewski:1996,
  title={Mean-field description of ground-state properties of drip-line nuclei: Pairing and continuum effects},
  author={Dobaczewski, J and Nazarewicz, W and Werner, TR and Berger, JF and Chinn, CR and Decharg{\'e}, J},
  journal={Physical Review C},
  volume={53},
  number={6},
  pages={2809},
  year={1996},
  publisher={APS}
}

@article{sorlin:2008,
    author = {Sorlin, O. and Porquet, M.-G.},
    title = {Nuclear magic numbers: New features far from stability},
    journal = {Progress in Particle and Nuclear Physics},
    volume = {61},
    issue = {2},
    pages = {602-673},
    year = {2008}
}

@article{orlandi:2018,
  title={{Neutron-hole states in $^{131}$Sn and spin-orbit splitting in neutron-rich nuclei}},
  author={Orlandi, R and Pain, SD and Ahn, S and Jungclaus, A and Schmitt, KT and Bardayan, DW and Catford, WN and Chapman, R and Chipps, KA and Cizewski, JA and others},
  journal={Physics Letters B},
  volume={785},
  pages={615--620},
  year={2018},
  publisher={Elsevier}
}

@article{beck:2021,
  title = {{Mass Measurements of Neutron-Deficient Yb Isotopes and Nuclear Structure at the Extreme Proton-Rich Side of the $N=82$ Shell}},
  author = {Beck, S\"onke and Kootte, Brian and Dedes, Irene and Dickel, Timo and Kwiatkowski, A. A. and Lykiardopoulou, Eleni Marina and Pla\ss{}, Wolfgang R. and Reiter, Moritz P. and Andreoiu, Corina and Bergmann, Julian and Brunner, Thomas and Curien, Dominique and Dilling, Jens and Dudek, Jerzy and Dunling, Eleanor and Flowerdew, Jake and Gaamouci, Abdelghafar and Graham, Leigh and Gwinner, Gerald and Jacobs, Andrew and Klawitter, Renee and Lan, Yang and Leistenschneider, Erich and Minkov, Nikolay and Monier, Victor and Mukul, Ish and Paul, Stefan F. and Scheidenberger, Christoph and Thompson, Robert I. and Tracy, James L. and Vansteenkiste, Michael and Wang, Hua-Lei and Wieser, Michael E. and Will, Christian and Yang, Jie},
  journal = {Phys. Rev. Lett.},
  volume = {127},
  issue = {11},
  pages = {112501},
  numpages = {8},
  year = {2021},
  month = {Sep},
  publisher = {American Physical Society},
  doi = {10.1103/PhysRevLett.127.112501},
  url = {https://link.aps.org/doi/10.1103/PhysRevLett.127.112501}
}

@article{jesch:2015,
    author = {Christian Jesch and Timo Dickel and Wolfgang R. Plaß and Devin Short and Samuel Ayet San Andres and Jens Dilling and Hans Geissel and Florian Greiner and Johannes Lang and Kyle G. Leach and Wayne Lippert and Christoph Scheidenberger and Mikhail I. Yavor },
    title = {{The MR-TOF-MS isobar separator for the TITAN facility at TRIUMF}},
    journal = {TCP 2014},
    pages = {175-184},
    year = {2017}
}

@article{spataru:2025,
  title = {Broadband mass measurements with the {FRS} Ion Catcher facility at {GSI} and theory developments investigating the shape-phase transition near ${N}=90$},
  author = {Sp\ifmmode \u{a}\else \u{a}\fi{}taru, Anamaria and Kripk\'o-Koncz, Gabriella and Dickel, Timo and Hornung, Christine and Pla\ss{}, Wolfgang R. and Constantin, Paul and Amanbayev, Daler and Ayet San Andr\'es, Samuel and Balabanski, Dimiter L. and Beck, Soenke and Bergmann, Julian and Geissel, Hans and Kalantar-Nayestanaki, Nasser and Kehat, Jonathan and Mardor, Israel and Minkov, Nikolay and Mollaebrahimi, Ali and Scheidenberger, Christoph and Wasserhe\ss{}, Max and Wilsenach, Heinrich and Zhao, Jianwei},
  journal = {Phys. Rev. C},
  volume = {111},
  issue = {5},
  pages = {054307},
  numpages = {13},
  year = {2025},
  month = {May},
  publisher = {American Physical Society},
  doi = {10.1103/PhysRevC.111.054307},
  url = {https://link.aps.org/doi/10.1103/PhysRevC.111.054307}
}

@article{liu:2026,
    author = {B. Liu and M. Brodeur and J.A. Clark and D. Ray and G. Savard and A.A. Valverde and D.P. Burdette and A.M. Houff and A. Mitra and G.E. Morgan and R. Orford and W.S. Porter and C. Quick and F. Rivero and K.S. Sharma and L. Varriano},
    title = {Precise Mass Measurement of the $^{149}${La}-$^{149}${Ce}-$^{149}${Pr} isobaric chain},
    journal = {arXiv preprint arXiv 2601.09959},
    year = {2026},
}

@article{kimura:2024,
  title = {Comprehensive mass measurement study of $^{252}\mathrm{Cf}$ fission fragments with {MRTOF-MS} and detailed study of masses of neutron-rich {Ce} isotopes},
  author = {Kimura, S. and Wada, M. and Haba, H. and H. Ishiyama and Ishizawa, S. and Ito, Y. and Niwase, T. and Rosenbusch, M. and Schury, P. and Takamine, A.},
  journal = {Phys. Rev. C},
  volume = {110},
  issue = {4},
  pages = {045810},
  numpages = {9},
  year = {2024},
  month = {Oct},
  publisher = {American Physical Society},
  doi = {10.1103/PhysRevC.110.045810},
  url = {https://link.aps.org/doi/10.1103/PhysRevC.110.045810}
}

@article{mgm:1948,
  title = {On Closed Shells in Nuclei},
  author = {Mayer, Maria G.},
  journal = {Phys. Rev.},
  volume = {74},
  issue = {3},
  pages = {235--239},
  numpages = {0},
  year = {1948},
  month = {Aug},
  publisher = {American Physical Society},
  doi = {10.1103/PhysRev.74.235},
  url = {https://link.aps.org/doi/10.1103/PhysRev.74.235}
}

@article{mgm:1949,
  title = {On Closed Shells in Nuclei. II},
  author = {Mayer, Maria Goeppert},
  journal = {Phys. Rev.},
  volume = {75},
  issue = {12},
  pages = {1969--1970},
  numpages = {0},
  year = {1949},
  month = {Jun},
  publisher = {American Physical Society},
  doi = {10.1103/PhysRev.75.1969},
  url = {https://link.aps.org/doi/10.1103/PhysRev.75.1969}
}

@article{haxel:1949,
  title = {On the "Magic Numbers" in Nuclear Structure},
  author = {Haxel, Otto and Jensen, J. Hans D. and Suess, Hans E.},
  journal = {Phys. Rev.},
  volume = {75},
  issue = {11},
  pages = {1766},
  numpages = {0},
  year = {1949},
  month = {Jun},
  publisher = {American Physical Society},
  doi = {10.1103/PhysRev.75.1766.2},
  url = {https://link.aps.org/doi/10.1103/PhysRev.75.1766.2}
}

@article{brown:2001,
title = {The nuclear shell model towards the drip lines},
journal = {Progress in Particle and Nuclear Physics},
volume = {47},
number = {2},
pages = {517-599},
year = {2001},
issn = {0146-6410},
doi = {https://doi.org/10.1016/S0146-6410(01)00159-4},
url = {https://www.sciencedirect.com/science/article/pii/S0146641001001594},
author = {B.A. Brown}
}

@article{bohr:1958,
  title = {Possible Analogy between the Excitation Spectra of Nuclei and Those of the Superconducting Metallic State},
  author = {Bohr, A. and Mottelson, B. R. and Pines, D.},
  journal = {Phys. Rev.},
  volume = {110},
  issue = {4},
  pages = {936--938},
  numpages = {0},
  year = {1958},
  month = {May},
  publisher = {American Physical Society},
  doi = {10.1103/PhysRev.110.936},
  url = {https://link.aps.org/doi/10.1103/PhysRev.110.936}
}

@misc{bohr:book,
  title={Nuclear structure, vol. 1},
  author={Bohr, Aage and Mottelson, Ben R.},
  year={1969},
  publisher={W. A. Benjamin, Inc.}
}

@misc{ring:book,
  title={The Nuclear Many-Body Problem},
  author={Ring, Peter and Schuck, Peter and Strayer, MR},
  year={1983},
  publisher={American Institute of Physics}
}

@article{huang2021ame2020i,
  title   = {{The AME 2020 atomic mass evaluation (I). Evaluation of input data, and adjustment procedures}},
  author  = {Huang, W. J. and Wang, Meng and Kondev, F. G. and Audi, G. and Naimi, S.},
  journal = {Chinese Physics C},
  volume  = {45},
  number  = {3},
  pages   = {030002},
  year    = {2021},
  doi     = {10.1088/1674-1137/abddaf}
}
\end{document}


\title{Supplemental Material for: Interplay between Nuclear Shell Structure and Pairing around Doubly Magic $^{132}$Sn}

\author{R.~Simpson\text{\orcidicon{0009-0002-7714-5649}}}
\email{ranes@triumf.ca}
\affiliation{TRIUMF, Vancouver, British Columbia, V6T 2A3, Canada}
\affiliation{Department of Physics \& Astronomy, University of British Columbia, Vancouver, British Columbia V6T 1Z1, Canada}

\author{G.~Palkanoglou\text{\orcidicon{0000-0001-8543-6097}}}
\email{gpalkanoglou@triumf.ca}
\affiliation{TRIUMF, Vancouver, British Columbia, V6T 2A3, Canada}

\author{E.C.~Brisley}
\affiliation{TRIUMF, Vancouver, British Columbia, V6T 2A3, Canada}
\affiliation{Department of Physics \& Astronomy, University of British Columbia, Vancouver, British Columbia V6T 1Z1, Canada}

\author{J.D.~Cardona}
\affiliation{TRIUMF, Vancouver, British Columbia, V6T 2A3, Canada}
\affiliation{Department of Physics and Astronomy, University of Manitoba, Winnipeg, Manitoba, R3T 2N2 Canada}

\author{A.~Czihaly}
\affiliation{TRIUMF, Vancouver, British Columbia, V6T 2A3, Canada}
\affiliation{Department of Physics and Astronomy, University of Victoria, Victoria, British Columbia, V8O 5C2, Canada}

\author{S.~Kakkar}
\affiliation{TRIUMF, Vancouver, British Columbia, V6T 2A3, Canada}
\affiliation{Department of Physics and Astronomy, University of Manitoba, Winnipeg, Manitoba, R3T 2N2 Canada}

\author{M.~Simonov}
\affiliation{II. Physikalisches Institut, Justus-Liebig-Universität, Gießen, 35392, Germany.}
\affiliation{Helmholtz Forschungsakademie Hessen für FAIR (HFHF), Campus Gießen, Gießen, 35392, Germany.}

\author{E.~Taylor}
\affiliation{TRIUMF, Vancouver, British Columbia, V6T 2A3, Canada}
\affiliation{Department of Physics and Astronomy, University of Western Ontario, London, Ontario, N6A 3K7 Canada}

\author{C.~Walls}
\affiliation{TRIUMF, Vancouver, British Columbia, V6T 2A3, Canada}
\affiliation{Department of Physics and Astronomy, University of Manitoba, Winnipeg, Manitoba, R3T 2N2 Canada}

\author{P.~Weligampola}
\affiliation{TRIUMF, Vancouver, British Columbia, V6T 2A3, Canada}
\affiliation{Department of Physics and Astronomy, University of Manitoba, Winnipeg, Manitoba, R3T 2N2 Canada}

\author{C.~Chambers}
\affiliation{TRIUMF, Vancouver, British Columbia, V6T 2A3, Canada}

\author{F.~Maldonado Mill\'{a}n}
\affiliation{TRIUMF, Vancouver, British Columbia, V6T 2A3, Canada}

\author{A.~Mollaebrahimi}
\affiliation{II. Physikalisches Institut, Justus-Liebig-Universität, Gießen, 35392, Germany.}
\affiliation{GSI Helmholtz Center for Heavy Ion Research, Campus Gie\ss en, 35392, Gie\ss en, Germany}

\author{D.~Ray}
\affiliation{TRIUMF, Vancouver, British Columbia, V6T 2A3, Canada}

\author{A.~Weaver}
\affiliation{TRIUMF, Vancouver, British Columbia, V6T 2A3, Canada}

\author{J.~Yu}
\affiliation{GSI Helmholtz Center for Heavy Ion Research, Campus Gie\ss en, 35392, Gie\ss en, Germany}

\author{I.~Dillmann}
\affiliation{TRIUMF, Vancouver, British Columbia, V6T 2A3, Canada}
\affiliation{Department of Physics and Astronomy, University of Victoria, Victoria, British Columbia, V8O 5C2, Canada}

\author{A.~Gezerlis\text{\orcidicon{0000-0003-2232-2484}}}
\email{gezerlis@uoguelph.ca}
\affiliation{Department of Physics, University of Guelph, Guelph, ON N1G 2W1, Canada}

\author{G.~Gwinner}
\affiliation{TRIUMF, Vancouver, British Columbia, V6T 2A3, Canada}
\affiliation{Department of Physics and Astronomy, University of Manitoba, Winnipeg, Manitoba, R3T 2N2 Canada}

\author{A.~O.~Macchiavelli}
\affiliation{Physics Division, Oak Ridge National Laboratory, Oak Ridge, Tennessee 37831, USA}

 \author{S.~Malbrunot-Ettenauer\text{\orcidicon{0009-0004-3323-6500}}}
\email{sette@triumf.ca}
\affiliation{TRIUMF, Vancouver, British Columbia, V6T 2A3, Canada}
\affiliation{Department of Physics, University of Toronto, Toronto, Ontario, M5S 1A7, Canada}

\author{M.P.~Reiter}
\affiliation{School of Physics and Astronomy, The University of Edinburgh, Edinburgh, EH9 3FD, Scotland, UK}

\author{A. A. Kwiatkowski}
\affiliation{TRIUMF, Vancouver, British Columbia, V6T 2A3, Canada}
\affiliation{Department of Physics and Astronomy, University of Victoria, Victoria, British Columbia, V8O 5C2, Canada}

\maketitle 
In this supplemental material document we provide additional details quantifying the arguments made in the main text. We present calculations of the pairing correlation energies, the underlying single-particle states employed, as well as, some details on our deformed Hartree-Fock-Bogolyubov (HFB) formulation. Finally we present the neutron odd-even staggering (OES) on isotones away from the $N=82$ shell closure demonstrating the contrast with the mid-shell behavior.

\begin{figure}[t]
    \centering
    \includegraphics[width=0.8\columnwidth]{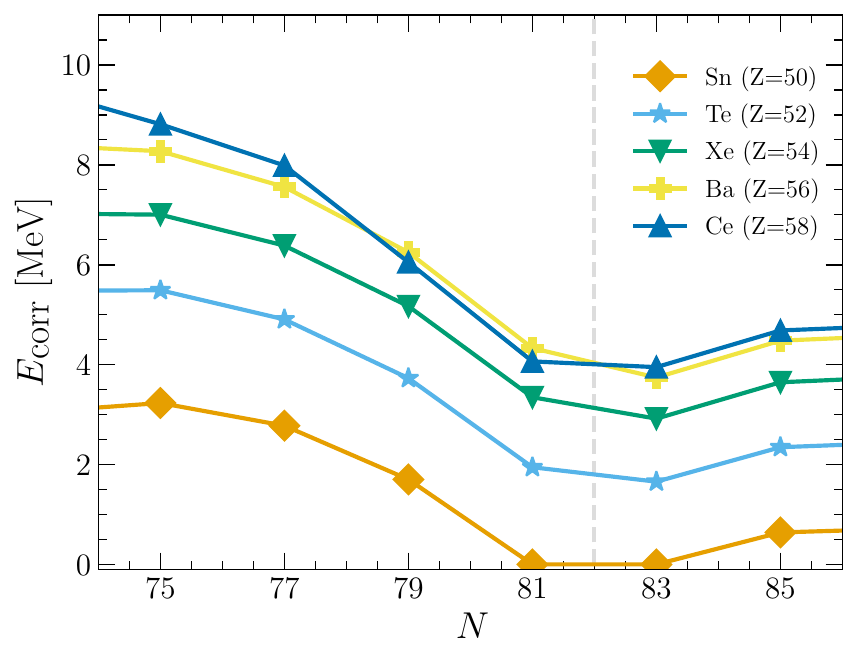}
    \caption{The pairing correlation energy of even-$Z$ isotopes, which acts as the order parameter of the superfluid state at the ground state of the nuclei, quantifying their pairing correlations.}
    \label{fig:ecorr}
    \end{figure}

\PRLsection{Pairing correlation energy from HFB}
In the main text we performed HFB calculations of the OES using 
\begin{align}
    \Delta^{(3)}(n) = \frac{(-1)^n}{2}\left[B(n+1)-2B(n)+B(n-1)\right]~, \label{eq:oes}
\end{align}
and replacing the binding energy with the pairing correlation energy defined as,
\begin{align}
    E_{\textrm{corr}}(N,Z)=E_{\textrm{HFB}}(N,Z)-E_{\textrm{NS}}(N,Z)~,
\end{align}
where $E_{\textrm{NS}}(N,Z)$ stands for the energy of a normal (non-superfluid) state. Since HFB theory is variational, $E_{\textrm{corr}}$ quantifies the strength of the pairing correlations, with vanishing $E_\textrm{corr}$ signifying a normal-state ground state. This means that $E_\textrm{corr}$ can act as an order parameter for the superfluid phase in the nucleus' ground state. In extended systems this role can also be played by the OES but in systems with finite size an increase in $E_\textrm{corr}$ might not translate to larger OES due to shell-effects: this is at the center of the arguments put forth in the main text. Indeed, the pairing correlations along the $N=81$ and $N=83$ isotones grow as one moves away from $Z=50$ as shown in Fig.~\ref{fig:ecorr} of this supplemental material. Along the $N=81$ and $N=83$ isotones, the correlation energy grows with $Z$, reaching a plateau at $Z=58$. Comparing with the trends seen in Fig.~1 (c)-(e) of the main text, the picture described there emerges: the pairing correlation energy grows with the proton number, antagonizing the shell-effects' reduction of the OES across shells. 

\begin{figure}[t]
    \centering
    \includegraphics[width=1\columnwidth]{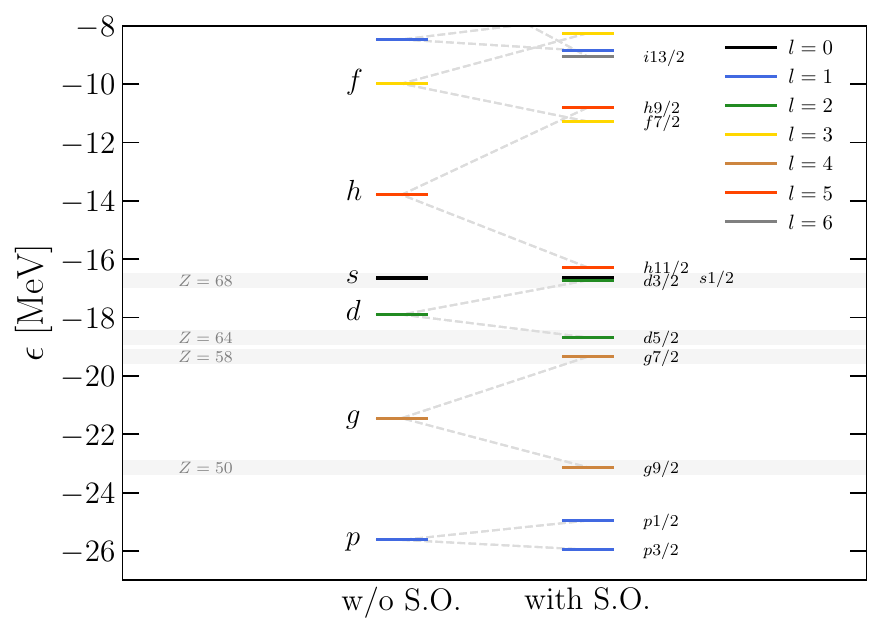}
    \caption{A single-particle spectrum employed in the HFB calculations. The orbital angular momentum and the total angular momentum (when employing spin-orbit splitting) are also shown for each state. The shell-closures are marked by gray bands. The single-particle spectrum is $N,Z$-dependent [see Eq.~(\ref{eq:ws})] and demonstrated here for $N=Z=82$; we find that the features described in the main text are independent of $Z$.}
    \label{fig:spstates}
    \end{figure}

\PRLsection{The shell and sub-shell structure}
In the main text we have repeatedly mentioned the expected shell structure around $^{132}$Sn. In Fig.~\ref{fig:spstates} of this supplemental material we present the single-particle states employed in our calculations which dictate the shell structure underlying our HFB description. These states are obtained by solving a Schr{\"o}dinger's equation with a Wood-Saxon trapping potential and with or without spin-orbit splitting. The (sub)shell closures are marked by gray bands. The shell gaps shown in Fig.~\ref{fig:spstates} of this supplemental material are essential in reproducing the experimental trends of Fig.~1 (c)-(e) in the main text. Consequently, we find that the spin-orbit splitting that dictates them is equally important. 

\PRLsection{Deformed HFB calculations} Our deformed HFB calculations treat deformation statically by deforming the Wood-Saxon trapping potential that generates our single-particle states:
\begin{align}
    V_{\textrm{WS}}(\mathbf{r};N,Z,\boldsymbol{\beta}) = V_0(N,Z) \left[1+e^{l(\mathbf{r};\boldsymbol{\beta})/a}\right]^{-1}~, \label{eq:ws}
\end{align}
where $V_0(N,Z)$ contains an $N-Z$ dependence~\cite{bohr:book,orlandi:2018}, $l(\mathbf{r})$ is the length from the deformed nuclear surface,  and $a=0.67~\textrm{fm}$ is the surface's diffuseness; in our spherical formulation the $\boldsymbol{\beta}$-dependence is dropped. For more details on our deformed HFB formulation see Refs.~\cite{palkanoglou:2025a,palkanoglou:2025b}. We do not determine the deformation parameters self-consistently but rather prescribe them based on their realistic predictions for the region. We have used the tabulation of Ref.~\cite{moller:2016} to guide our choices. For the calculations presented in Fig.\,1(e) in the main text, the nuclei were taken to be of quadrupole deformation independent of $Z$, i.e., $\beta_2(N)$, with $\beta_2(80)=\beta_2(81)=0.032$, $\beta_2(82)=\beta_2(84)=0$, and $\beta_2(83)=-0.032$. These values correspond to the average values expected for the relevant parts of these isotonic chains as predicted in Ref.~\cite{moller:2016}. 

The resulting OES provides a good description of the experimental data for $Z<60$ in the $N=83$ isotone, while the $N=81$ isotone reproduces the general trends but with diminished values.

For $Z>60$, the OES decreases to approximately $\sim0.4,\mathrm{MeV}$ at $Z=64$ in both isotones before increasing again, not reflecting the experimental trend in this region. This behavior originates from the non-self-consistent treatment of deformation in our model. Within the present approach, the underlying nuclear structure then evolves too rapidly from nucleus to nucleus for $Z>60$, causing $\Delta^{(3)}$ values derived from $E_{\mathrm{corr}}$ to no longer provide a reliable estimate of the experimentally observed OES. These findings motivate future work aimed at implementing a fully self-consistent treatment of deformation.

\begin{figure}[h]
    \centering
    \includegraphics[width=1\columnwidth]{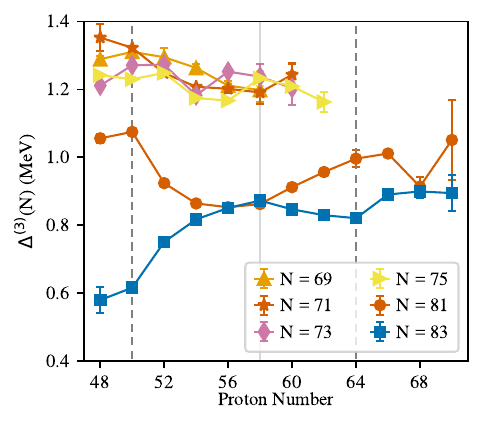}
    \caption{Odd-even staggering of binding energies: The $N = 81$ and  $N=83$ isotonic chains from Fig. 1(c) in the main text are pictured alongside analogous data from the odd-$N$ chains for $69 \leq N \leq 75$.}
    \label{fig:N69to75}
\end{figure}

\PRLsection{Available experimental data} Extending upon Fig.~1(c) from the main text, in Fig.~\ref{fig:N69to75} of this supplemental material, we display results for isotonic chains of odd-$N$ values from $N=69$ to $N = 75$ (all data taken from Ref.\,\cite{huang2021ame2020i}), in addition to the data shown in the main text. See main text for detailed discussion. 

\clearpage

\bibliography{sup_bibliography}